\documentstyle[preprint,tighten,aps]{revtex}

\newcommand{\lagrange}{{\cal L}}

\begin{document}


\draft
\preprint{DAMTP-98-11, hep-th/9802140}
\date{February 1998}
\title{N = 2 Supersymmetric Quantum Black Holes \\
       in Five Dimensional Heterotic String Vacua}
\author{Ingo Gaida}

\address{
Department of Applied Mathematics and Theoretical Physics \\
University of Cambridge, Silver Street, Cambridge CB3 9EW, UK}
\maketitle\

\begin{abstract}
Exact black hole solutions of the 
five dimensional heterotic $S$-$T$-$U$ model including all
perturbative quantum corrections and preserving $1/2$ of 
$N=2$ supersymmetry are studied.
It is shown that the quantum corrections yield a bound
on electric charges and harmonic functions of the solutions.
\end{abstract}



%
%


\newpage


\setcounter{page}{1}
\setcounter{section}{1}

In \cite{Strominger} Strominger and Vafa considered five dimensional
string theory with $N=4$ supersymmetry to derive the 
Bekenstein-Hawking entropy \cite{hawking} by counting black hole
microstates. In this letter the low-energy effective action
of the five dimensional $S$-$T$-$U$ model in heterotic string vacua
with $N=2$ supersymmetry is studied. This model yields the
Strominger-Vafa black hole including, in addition,
perturbative quantum corrections.
\\
The action of five dimensional $N=2$ supergravity coupled to 
$N=2$ vector multiplets has been constructed
in \cite{Sierra} and the compactification of $N=1$, $D=11$ supergravity
[M-theory]
down to five dimensionens on Calabi-Yau 3-folds ($CY_{3}$) with
Hodge numbers ($h_{1,1},h_{2,1}$) and topological intersection numbers
$C_{\Lambda\Sigma\Delta}$ has been given in \cite{Comp,Antoniadis}:
The $N_V$-dimensional space $\cal M$ ($N_V=h_{1,1}-1$) of scalar 
components of $N=2$ abelian vector multiplets coupled to supergravity
can be regarded as a hypersurface of a $h_{1,1}$-dimensional manifold
whose coordinates $X(\phi)$ are in correspondence with the vector bosons
(including the graviphoton). The definining equation of the hypersurface is 
${\cal V}(X)=1$ and the prepotential ${\cal V}$ 
is a homogeneous cubic polynomial
in the coordinates $X(\phi)$:
\begin{eqnarray}
{\cal V}(X) &=& \frac{1}{6} \ C_{\Lambda\Sigma\Delta}
                X^{\Lambda}X^{\Sigma}X^{\Delta},
\hspace{1cm} \Lambda,\Sigma,\Delta = 1, \ldots h_{1,1} 
\end{eqnarray}
In five dimensions the $N=2$ vector multiplet has a single scalar and
$\cal M$ is therefore real. Moreover, if the prepotential is 
factorizable, it is generically symmetric and of the form
\begin{eqnarray}
{\cal V}(X) &=& X^1 \ Q(X^{\Lambda+1}),
\hspace{1cm} \Lambda = 1, \ldots N_V 
\end{eqnarray}
where $Q$ denotes a quadratic form. It follows that the scalar 
fields parametrize the coset space
\begin{eqnarray}
\label{space}
 {\cal M} &=& SO(1,1) \ \times \ \frac{SO(N_V-1,1)}{SO(N_V-1)}. 
\end{eqnarray}
The bosonic action of  $N=2$ supergravity  coupled to 
$N_V$ vector multiplets is given by (omitting Lorentz indices)
\begin{eqnarray}
e^{-1} \lagrange &=& - \frac{1}{2} R
                     - \frac{1}{2} g_{ij} \partial\phi^i \partial\phi^j
                     - \frac{1}{4} G_{\Lambda\Sigma} 
                          F^\Lambda F^\Sigma
                     + \frac{e^{-1}}{48} C_{\Lambda\Sigma\Delta} 
                          \epsilon F^\Lambda F^\Sigma A^\Delta.
\end{eqnarray}
The corresponding vector and scalar metrics are encoded in the function
${\cal V}$ completely
\begin{eqnarray}
 G_{\Lambda\Sigma} &=& - \frac{1}{2} \partial_\Lambda \partial_\Sigma
                       \ln {\cal V}(X)_{|{\cal V}=1},
\\
 g_{ij} &=& G_{\Lambda\Sigma} \partial_i X^\Lambda(\phi)
                              \partial_j X^\Sigma (\phi)_{|{\cal V}=1}.
\end{eqnarray}
Here the derivatives in the scalar metric are with
respect to the $h_{1,1}$ coordinates $X^\Lambda(\phi)$ and the   
$h_{1,1}-1$ scalar fields $\phi^i$, respectively. 
It is useful to introduce special coordinates $t^\Lambda$ and their
duals $t_\Lambda$ \cite{cham} :
\begin{eqnarray}
 t^\Lambda(\phi) &=& 6^{-1/3} X^\Lambda(\phi)
\ = \  C^{\Lambda\Sigma}(\phi)  t_\Sigma(\phi),
\nonumber\\
 t_\Lambda(\phi) &=& C_{\Lambda\Sigma\Delta} t^\Sigma(\phi) t^\Delta(\phi)
                 \ = \ C_{\Lambda\Sigma}(\phi) t^\Sigma(\phi)  
\end{eqnarray}
From these definitions follows
$t^\Lambda t_\Lambda = 1$ and
$C_{\Lambda\Sigma}  C^{\Sigma\Delta}=\delta_{\Lambda}^{ \ \Delta}$.  
In these special coordinates one finds for the gauge coupling matrix
\begin{eqnarray}
 G_{\Lambda\Sigma} &=& - \frac{6^{1/3}}{2} 
(C_{\Lambda\Sigma}- \frac{3}{2} t_\Lambda t_\Sigma),
\hspace{1cm}
  G^{\Lambda\Sigma}  \ = \ - \frac{2}{6^{1/3}} 
(C^{\Lambda\Sigma}- 3 t^\Lambda t^\Sigma)
\end{eqnarray}
with $ G_{\Lambda\Sigma}  G^{\Sigma\Delta}  \ = \ \delta_{\Lambda}^{ \ \Delta}$
and 
$g_{ij} = -3 C_{\Lambda\Sigma} \partial_i t^\Lambda \partial_j t^\Sigma$. 
\\
It has been shown in \cite{FerKal1,cham} that the supersymmetry
transformations of the gaugino and the gravitino vanish if
the (electric) central charge $Z=t^\Lambda q_\Lambda$,
appearing in the supersymmetry algebra, 
has been minimized in moduli space 
($\partial_i Z = 0$). This minimization procedure
yields the fixed values of the
moduli on the black hole horizon \cite{FerKal1,la/wi}.
Equivalently one may use the ``stabilisation equations'' 
\begin{eqnarray}
q_\Lambda &=& t_\Lambda \  Z_{\rm fix},
\hspace{1cm}
Z_{\rm fix}^2 \ = \ C^{\Lambda\Sigma}_{\rm fix} q_\Lambda q_\Sigma.
\end{eqnarray}
The geometry of the corresponding
extreme $D=5$ black holes is determined
by the following metric 
\begin{eqnarray}
\label{metric}
ds^2 &=& - e^{-4V(r)} \ dt^2 \ + \ e^{2V(r)} (dr^2 + r^2 \ d\Omega_3^2)
\end{eqnarray}
where the metric function $e^{2V(r)}$ is a function
of harmonic functions. The moduli for so-called double-extreme black holes
are constant and given by their fixed values throughout the entire
space-time \cite{shmakova}. 
For these double-extreme black holes the gauge fields satisfy 
$2 \sqrt{-g}G_{\Lambda\Sigma}F^{\Sigma}=q_\Lambda$. Moreover,
the entropy \cite{hawking} of extreme black holes in five dimensions
is given by \cite{FerKal1}
\begin{eqnarray}
S_{BH} &=& \frac{A}{4G_N} \ = \ 
           \frac{\pi^2}{2 G_N}\ |\frac{Z}{3}|_{|{\rm fix}}^{3/2}.
\end{eqnarray}
In $D=5$ point-like objects are dual to string-like objects. 
Thus, corresponding to the electric central charge $Z$
exist the dual magnetic central charge $Z_m=t_\Lambda p^\Lambda$
with magnetic charges $p^\Lambda$. The electric and magnetic charges
arise in M-theory from two- and five-brane solitons which wrap even cycles
in the CY-space \cite{Comp,Becker}. 
\begin{eqnarray}
q_\Lambda &=& \int_{C^{4\Lambda} \times S_3}
              G_7,
\hspace{2cm}
p^\Lambda \ = \ \int_{C_2^\Lambda \times S_2}
              F_4. 
\end{eqnarray}
Here, $F_4$ is the field-strength of the three-form in $D=11$ supergravity
while $G_7= \frac{\delta\lagrange}{\delta F_4}$ is its dual; 
$C^{4\Lambda}$ [$C_2^\Lambda$] denotes
a four- [two-] cycle in $CY_3$. 
From the point of view of the heterotic string $q_{2,3}$ correspond to
perturbative electric charges of Kaluza-Klein excitations and winding modes,
$p^1$ is the charge of the fundamental string and $p^{2,3}$ [$q_1$] 
arise from $D=10$ solitonic five-branes wrapping around 
$K_3$ [$K_3 \times S_1$].
The magnetic central charge $Z_m$ determines the tension of 
magnetic string states as a function of the moduli. Thus,
analogous to the fixed value of the electric central charge,
there exist a fixed value for the string tension \cite{CBG,rahmfeld}.
\begin{eqnarray}
p^\Lambda &=& t^\Lambda \  Z_{m, \rm fix},
\hspace{1cm}
Z_{m, \rm fix}^3 \ = \ 27 C_{\Lambda\Sigma\Delta} p^\Lambda p^\Sigma p^\Delta.
\end{eqnarray}
It follows that the $D=5$ entropy-density of the magnetic string   is
given by \cite{CBG}
\begin{eqnarray}
S_{S} &\sim& |Z_m|_{|{\rm fix}}^{2} \sim 
\left (
 C_{\Lambda\Sigma\Delta} p^\Lambda p^\Sigma p^\Delta
\right )^{2/3}.
\end{eqnarray}
Compactifying the $D=10$ effective heterotic string  
on $K3 \times S_{1}$ one can construct the $D=5$, $N=2$
$S$-$T$-$U$ model \cite{ka}. This model contains
244 neutral hypermultiplets, which we will ignore in the following.
Moreover it contains three vector moduli $S$, $T$ and $U$, where $S$ denotes
the heterotic dilaton and $T,U$ are associated
to the graviphoton and the additional $U(1)$ gauge boson of
the $S_{1}$ compactification.
The $D=5$ heterotic $S$-$T$-$U$ model is dual to M-theory compactified
on a Calabi-Yau threefold \cite{Comp}. 
Further compactification on $S_{1}$
yields the rank 4 $S$-$T$-$U$ model in $D=4$, which
is dual to the $X_{24}(1,1,2,8,12)$ model of the
type II string compactified on a Calabi-Yau 
\cite{ka}.   
In special coordinates the prepotential reads
\begin{eqnarray}
\label{pre}
 {\cal V}(S,T,U)  &=& STU \ + \ h(T,U)
\end{eqnarray}
The function
$h(T,U)$ denotes perturbative quantum corrections, which have been
determined in  \cite{Antoniadis} 
\begin{eqnarray}
h(T,U) &=& \frac{a}{3} \ U^3 \ \theta (T-U) \ + \
           \frac{a}{3} \ T^3 \ \theta (U-T).
\end{eqnarray}
Here we have introduced the parameter $a=1$ in order to discuss
the classical limit $a \rightarrow 0$ in the following explicitly.
In the classical limit the scalar fields parametrize the coset
(\ref{space}) with $N_V=2$. 
Using very special geometry the dilaton field $S$ can be eliminated
through the algebraic equation
\begin{eqnarray}
S &=& \frac{1-h(T,U)}{TU}.
\end{eqnarray}
For convenience we define the functions
\begin{eqnarray}
f(x,y) &=& \frac{2a}{3} \ x^3 \ \theta (y-x) 
           - \frac{a}{3} \ y^3 \ \theta (x-y),
\nonumber\\
g(x,y) &=& \frac{a}{3} \ x^3 \ \delta (y-x) 
           - \frac{a}{3} \ y^3 \ \delta (x-y).
\end{eqnarray}
It follows
\begin{eqnarray}
\partial_T S &=& - \frac{1+f(T,U)}{T^2U} - \frac{g(U,T)}{TU},
\nonumber\\
\partial_U S &=& - \frac{1+f(U,T)}{U^2T} - \frac{g(T,U)}{TU}.
\end{eqnarray}
If we take $t^{1,2,3}=(S,T,U)$, we find for the dual coordinates
\begin{eqnarray}
t_1 &=& \frac{1}{3} TU,
\nonumber\\
t_2 &=& \frac{1}{3} SU + \frac{a}{3} T^2 \theta (U-T)
\nonumber\\
t_3 &=& \frac{1}{3} ST + \frac{a}{3} U^2 \theta (T-U)
\end{eqnarray}
Thus, for the matrix $C$ 
(with components $C_{\Lambda\Sigma}$)
we obtain
\begin{eqnarray}
C &=& \frac{1}{6}
\left (
   \begin{array}{ccc}
   0     & U                        & T       \\
   U     & \ 2a T \theta (U-T) \  &  \ \frac{1-h(T,U)}{TU} \  \\
   T     & \ \frac{1-h(T,U)}{TU} \  & \ 2a U \theta (T-U)   \\
\end{array}
\right ).
\end{eqnarray}
Hence, the gauge coupling matrix reads
\begin{equation}
G = \frac{1}{2 \cdot 6^{2/3}}
\left (
\begin{array}{ccc}
T^2 U^2   & U  f(T,U)                            & T  f(U,T) \\
U f(T,U)&  \ \frac{1}{T^2} [1-2h(T,U)+f^2(T,U)] \ 
                                           & 2h(T,U) \ \frac{1-h(T,U)}{TU} \\  
T f(U,T)&2h(T,U) \ \frac{1-h(T,U)}{TU}
                                     & \ \frac{1}{U^2}[1-2h(T,U)+f^2(U,T)] \\
\end{array}
\right ).
\end{equation}
Moreover, it is straightforward to compute the metric $g_{ij}$ 
of the scalar fields
\begin{equation}
g =
\left (
\begin{array}{cc}
\frac{1}{T^2}[1-h(T,U)+T g(U,T)]  &  
    \frac{1}{2TU} [1+2h(T,U)+Tg(U,T)+Ug(T,U)]  \\
\frac{1}{2TU} [1+2h(T,U)+Tg(U,T)+Ug(T,U)] & 
  \frac{1}{U^2} [1-h(T,U)+U g(T,U)]        
\end{array}
\right )
\end{equation}
It follows in the weak coupling regime $S>T>U>0$ 
\begin{eqnarray}
\det g &=& \frac{3}{4} \frac{1}{T^2 U^2} - a \frac{U}{T^2} \\
\det G &=& \frac{1}{288}
\left (
  1 - \frac{a}{3} U^3
\right )
 \  
\left (
1 - \frac{a}{3} U^3 - a^2 U^6 + \frac{a^3}{27} U^9 
\right ) 
\end{eqnarray}
Note that the gauge coupling matrix depends only on $U$.
Thus, one obtains for the boundaries of the Weyl-chamber
$S>T>U$
\begin{center}
\begin{tabular}{|c|c|c|}
\hline
\ boundary \ & $\det g$ & \ critical points \  \\
\hline
$U \rightarrow 0$ & diverges & -  \\
\hline
$S \rightarrow T $ &  regular & 
$U_{\rm crit.}= \left ( (\frac{3}{a})^{1/3}, (\frac{3}{4a})^{1/3} \right )$ \\
\hline
$S \rightarrow T \rightarrow U $ &  degenerates & - \\
\hline
$T \rightarrow U $ &  regular & 
        $U_{\rm crit.}=  (\frac{3}{4a})^{1/3} $ \\
\hline
\end{tabular}
\end{center}
Here the boundaries are regular up to the critical points with
$\det g_{\rm crit.} = (0,\infty)$. The chamber $S>T>U>0$ has 
three boundaries. The lines $S=T$  and $T=U$  are generically
regular. These two lines intersect at one point in moduli
space ($S=T=U$). Classically this intersection point is a 
``double self-dual point'', i.e. this point is self-dual with respect
to T-duality ($R=1$) and S-duality ($g_5=1$). Including quantum 
corrections one obtains
\begin{eqnarray}
 U_0 &=& (1+\frac{a}{3})^{-1/3} \ \equiv \ U_{\rm crit.} (S \rightarrow T)
\ \equiv \ U_{\rm crit.} (T \rightarrow U)
\end{eqnarray} 
at this point. Thus, the scalar metric
degenerates at this point and, therefore, the moduli space
simply ends here \cite{Witten_2}. 
\\
For convenience we will restrict ourselves now
to the fundamental Weyl chamber $T>U$. Moreover, we will consider
first of all double-extreme black hole solutions before studying
the bigger class of extreme solutions given in \cite{wafic_1}.
Starting with the prepotential (\ref{pre})
and the constraint ${\cal V}(X)=1$ one obtains\footnote{In this 
double-extreme context all the operators take their fixed values 
in moduli space.}
from the electric stabilisation equations
\begin{eqnarray}
 3q_1 \ = \ ZTU, \hspace{1cm}  
 3q_2 \ = \ ZSU, \hspace{1cm}  
 3q_3 \ = \ ZST + a ZU^2.
\end{eqnarray}
It follows
\begin{eqnarray}
 (2aU^3+3)Z-9q_3U &=& 0,
\nonumber\\
 aZ^2 U^4 - 3 q_3 Z U^2 + 9 q_1 q_2 &=& 0.
\end{eqnarray}
In the classical limit $(a=0)$ one obtains 
for the fixed values of the fields \cite{rahmfeld}
\begin{eqnarray}
 S   \ = \  \left(\frac{q_2 q_3}{q_1^2}\right)^{1/3}, \hspace{1cm}  
 T   \ = \  \left(\frac{q_1 q_3}{q_2^2}\right)^{1/3}, \hspace{1cm}  
 U   \ = \  \left(\frac{q_1 q_2}{q_3^2}\right)^{1/3}.
\end{eqnarray} 
and the central charge
$ Z=3 (q_1 q_2 q_3)^{1/3}$.
Thus, we obtain the Strominger-Vafa black hole \cite{Strominger}
with entropy
\begin{eqnarray}
\label{classical}
 S_{BH}  &=& \frac{\pi^2}{2 G_N} \ \sqrt{q_1 q_2 q_3}.
\end{eqnarray}
Including the quantum corrections $(a=1)$ one obtains a quadratic
equation in $U^3$ with solution
\begin{eqnarray}
 U^3 &=& - \gamma (1-\sqrt{1-\delta/\gamma^2}) 
\nonumber\\
 \gamma &=& \frac{3}{2a} 
            \left (
  \frac{4aq_1q_2-3q_3^2}{4aq_1q_2+3q_3^2}
            \right)
\nonumber\\
 \delta &=& \frac{9q_1q_2}{4a^2q_1q_2+3aq_3^2}
\end{eqnarray}
Since $U$ is real we obtain a bound $\gamma^2-\delta \geq 0$,
which becomes, in terms of the charges,
\begin{eqnarray}
\label{bound}
 q_3^2  &\geq& 4 q_1 q_2.
\end{eqnarray} 
The appearance of this bound is a true quantum effect. 
The corresponding fixed values of the moduli $S,T$ and the central charge
follow from the solution straightforward. Note that the solution also has
to satisfy the inequality $S>T>U$ in terms of the charges. In the
classical limit this condition is satisfied if $q_3>q_2>q_1$. It follows
$q_3^2>q_1 q_2$ and, therefore, the quantum bound is 
stronger\footnote{I thank M. Green for a discussion
on this point.}.
If we consider, for convenience, the case where (\ref{bound}) is saturated, 
we obtain for the fixed values of the fields
\begin{eqnarray}
S &=& \sqrt{\frac{q_2}{q_1}}  \left( \frac{3}{4} \right )^{1/3}
\hspace{2cm} 
T \ = \ \sqrt{\frac{q_1}{q_2}}  \left( \frac{3}{4} \right )^{1/3}
\hspace{2cm} 
U \ = \  \left( \frac{3}{4} \right )^{1/3}
\end{eqnarray} 
It follows that the black hole entropy is given by
\begin{eqnarray}
S_{BH} &=& \frac{\pi^2}{6 G_N} \  (q_3)^{3/2}.
\end{eqnarray}
Clearly this result does not coincide with the classical entropy
(\ref{classical}) in the limit $q_3^2 = 4 q_1 q_2$.
Note that the metric function is always given by
$e^{2V}=1+ \frac{Z}{r^3}$ in the double extreme limit \cite{wafic_1}.
Moreover, the entropy vanishes if one of the electric charges vanishes.
The dual string solution as been extensively discussed in the
literature \cite{be/mo,CBG,rahmfeld}. 
The fixed values of the scalar fields are given by
$S,T,U=p^{1,2,3}/Z_m$ and the fixed value
of the magnetic central charge reads
\begin{eqnarray}
 Z_{m} &=& 3 
\left ( 
 p^1 p^2 p^3 + \frac{a}{3} (p^3)^3
\right )^{1/3}.
\end{eqnarray} 
In the classical limit the electric and
magnetic central charge are dual to each other, if one exchanges
electric and magnetic charges. This property does not hold at the 
quantum level. It follows that some of the magnetic charges can vanish
to give a non-trivial entropy-density of the dual magnetic string.
\\
Now we will consider the more general class of black hole solutions
of \cite{wafic_1}.
The static, spherically symmetric 
BPS black hole solution of \cite{wafic_1} has metric (\ref{metric})
and 
\begin{eqnarray}
 2 G_{\Lambda\Sigma} F^\Sigma_{0m} &=& e^{-4V(r)} \partial_m
H_\Lambda, \hspace{1cm} n,m=1,2,3,4
\nonumber\\
\eta^{nm} \partial_n \partial_m H_\Lambda(r) &=& 0 
\hspace{1cm} \Rightarrow \hspace{1cm}
H_\Lambda \ = \ h_\Lambda + \frac{q_\Lambda}{r^2} 
\end{eqnarray}
Here the five-dimensional harmonic functions $H_\Lambda$ 
are characterized by the
electric charge $q_\Lambda$ of the three abelian gauge fields
(including the graviphoton) and the arbitrary
constants $h_\Lambda$. For special values of $h_\Lambda$ we obtain
the double-extreme solution discussed above.
Moreover, the solution satisfies
\begin{eqnarray}
\label{ebh}
\sqrt{-g} \ t_\Lambda &=& \frac{1}{3} \ H_\Lambda.
\end{eqnarray}
From (\ref{ebh}) follows  
\begin{eqnarray}
 e^{-2V} H_1 \ = \ TU, \hspace{1cm}  
 e^{-2V} H_2 \ = \ SU, \hspace{1cm}  
 e^{-2V} H_3 \ = \ ST + a U^2.
\end{eqnarray}
Thus, analogous to the double-extreme black hole solution we obtain
\begin{eqnarray}
 (2aU^3+3) \ e^{2V} - 3 H_3 U &=& 0,
\nonumber\\
 a e^{4V}  U^4 - H_3 e^{2V} U^2 + H_1 H_2 &=& 0.
\end{eqnarray}
In the classical limit $(a=0)$ one finds \cite{wafic_1}
\begin{eqnarray}
 S   \ = \  \left(\frac{H_2 H_3}{H_1^2}\right)^{1/3}, \hspace{1cm}  
 T   \ = \  \left(\frac{H_1 H_3}{H_2^2}\right)^{1/3}, \hspace{1cm}  
 U   \ = \  \left(\frac{H_1 H_2}{H_3^2}\right)^{1/3}.
\end{eqnarray}
Including the quantum corrections $(a=1)$ one obtains again a quadratic
equation in $U^3$ with solution
\begin{eqnarray}
 U^3 &=& - \gamma (1-\sqrt{1-\delta/\gamma^2}) 
\nonumber\\
 \gamma &=& \frac{3}{2a} 
            \left (
  \frac{4a H_1 H_2 -3 H_3^2}{4aH_1H_2+3H_3^2}
            \right)
\nonumber\\
 \delta &=& \frac{9H_1H_2}{4a^2H_1H_2+3aH_3^2}
\end{eqnarray}
Since $U$ is real we obtain the bound $\gamma^2-\delta \geq 0$.
If we take, for instance, $4H_1H_2+3H_3^2 > 0$ we obtain, in 
terms of the harmonic functions,
\begin{eqnarray}
 H_3^2  &\geq& 4 H_1 H_2.
\end{eqnarray} 
The corresponding values for the moduli $S,T$ and the metric function
$e^{2V}$ in terms of harmonic functions follow straightforward. 
Note that this black hole configuration exhibits a $\bf Z_2$ 
symmetry:
$H_\Lambda  \rightarrow e^{i n \pi} H_\Lambda$ for integer $n$.
The corresponding black hole entropy of this extreme black hole
solution is by definition the same as for the double-extreme solution.
Although we can compute now the full quantum solution,
i.e. the values of the moduli on the horizon, the entropy and
the metric, these expressions are not very illuminating for the exact
solution. Instead we
give here the first order quantum corrections to various quantities
to give a qualitative discussion, i.e. we omitt contribution of order
${\cal O}(a^2)$. 
The corresponding fixed values of the moduli on the horizon are

\begin{eqnarray}
 S_{| {\rm fix}} =  \left(\frac{q_2 q_3}{q_1^2}\right)^{1/3}
( 1 - \alpha ),
\hspace{0,5cm}
 T_{| {\rm fix}} =  \left(\frac{q_1 q_3}{q_2^2}\right)^{1/3}
( 1 - \alpha ),
\hspace{0,5cm}
 U_{| {\rm fix}} =  \left(\frac{q_1 q_2}{q_3^2}\right)^{1/3}
( 1 + 2 \alpha )
\end{eqnarray}
with $\alpha= \frac{2aq_1 q_2}{9q_3^2}$. It follows
for the central charge 
$
 Z_{| {\rm fix}} =  3 (q_1 q_2 q_3)^{1/3}
 \left (
  1 - \alpha
 \right )$.
The corresponding black hole entropy is 
\begin{eqnarray}
 S_{BH}  &=& \frac{\pi^2}{2 G_N} \ \sqrt{q_1 q_2 q_3}
 \left (
  1 - \frac{2}{3} \ \alpha 
 \right ).            
\end{eqnarray}
Moreover, the leading order correction for the metric function $e^{2V}$
is given by
\begin{eqnarray}
 e^{2V}  &=& 
 \left (
  H_1 H_2 H_3
 \right )^{1/3}
 \left (
  1 - \Delta
  \right ),
\hspace{1cm}
\Delta = \frac{2a}{9} \frac{H_1 H_2}{H_3^2},
\end{eqnarray}
Near the horizon ($r=0$) the metric becomes approximately
\begin{eqnarray}
 ds^2 &=& - \frac{r^4}{\lambda^2} \ dt^2 \ + \
                 \frac{\lambda^2}{r^2} \ dr^2 \ + \
                 \lambda^2 d \Omega_3^2,
\hspace{1cm} \lambda^2 = (q_1 q_2 q_3)^{1/3} (1-\alpha)           
\end{eqnarray}
It follows that the five-dimensional space-time manifold ${\cal M}_5$ is a 
product space near the horizon 
${\cal M}_{5} = AdS_2 \ \times \ S^3$
with symmetry group $SO(2,1) \times SO(3)$.
It is straightforward to obtain the leading order 
quantum correction to the ADM-mass of this extreme black hole.
Using diffeomorphism invariance the metric can always be brought into
the following form: 
\begin{eqnarray}
 ds^2 &=& -  
 \left ( 1 - \frac{8 G_N}{3 \pi} \frac{M_{ADM}}{r^2} + \cdots \right )
 \ dt^2 \ + \ \cdots
\end{eqnarray}
Introducing ``dressed charges'' $\hat q_\Lambda = q_\Lambda/h_\Lambda$
and expanding the metric function one obtains
\begin{eqnarray}
 M_{ADM}  &=& \frac{\pi}{4 G_N} 
\left \{
 \left ( 1 + \frac{a}{3} \frac{h_1 h_2}{h_3^2}
 \right ) \ \sum_{\Lambda=1,2,3} \ \hat q_\Lambda \ 
 - a \ \frac{h_1 h_2}{h_3^2} \ \hat q_3
\right \}.
\end{eqnarray}
In the classical limit we obtain the results of \cite{wafic_1}.
Moreover, we find that there are no leading order quantum
corrections to the ADM-mass if 
\begin{eqnarray}
  \frac{\hat q_1 + \hat q_2}{\hat q_3} &=& 2. 
\end{eqnarray}
In addition, the extreme black hole solution has vanishing ADM-mass
if 
\begin{eqnarray}
 \frac{h_3^2}{h_1 h_2} &=& 
           \frac{a}{3} 
           \frac{2 \hat q_3 - \hat q_1 - \hat q_2}
                {\hat q_1 + \hat q_2+ \hat q_3}. 
\end{eqnarray}
Although this result only holds to the leading order one expects
a similar condition for the massless black hole configuration including
all quantum corrections.
\\
\\
To conclude, exact black hole solutions preserving $1/2$ of $N=2$ 
supersymmetry in the five dimensional
$S$-$T$-$U$ model including all 
perturbative quantum corrections
have been studied. It has been shown that the 
quantum corrections yield a new bound
on electric charges and harmonic functions of the solutions.
The appearence of bounds of this kind in $N=2$ supersymmetric
models in five and four dimensions has been previously studied
in \cite{klaus,rahmfeld}.
It would be very interesting to find the corresponding statistical
mechanical interpretation of the black hole entropy analogous to
the analysis of Strominger and Vafa \cite{Strominger}
including this quantum bound.  

\bigskip  \bigskip

\noindent
{\bf Acknowledgments}  \medskip \newline
I would like to thank T. Mohaupt for discussions. 
This work is supported by DFG.


%
%

\renewcommand{\arraystretch}{1}

\newcommand{\NP}[3]{{ Nucl. Phys.} {\bf #1} {(19#2)} {#3}}
\newcommand{\PL}[3]{{ Phys. Lett.} {\bf #1} {(19#2)} {#3}}
\newcommand{\PRL}[3]{{ Phys. Rev. Lett.} {\bf #1} {(19#2)} {#3}}
\newcommand{\PR}[3]{{ Phys. Rev.} {\bf #1} {(19#2)} {#3}}
\newcommand{\IJ}[3]{{ Int. Jour. Mod. Phys.} {\bf #1} {(19#2)}
  {#3}}
\newcommand{\CMP}[3]{{ Comm. Math. Phys.} {\bf #1} {(19#2)} {#3}}
\newcommand{\PRp} [3]{{ Phys. Rep.} {\bf #1} {(19#2)} {#3}}

\end{document}